# Filling of a cavity with zero-point electromagnetic radiation

Jiří J. Mareš, V. Špička, J. Krištofik and P. Hubík

*Institute of Physics ASCR*

*Cukrovarnická 10, 162 53 Prague 6*

*Czech Republic*

**Abstract:** In the present contribution we analyse a simple thought process at $T = 0$ in an idealized heat engine having partitions made of a material with an upper frequency cut-off and bathed in zero-point (ZP) electromagnetic radiation. As a result, a possible mechanism of filling real cavities with ZP radiation based on Doppler's effect has been suggested and corresponding entropy changes are discussed.

## Introduction

Thought (Gedanken) experiments with idealized Carnot's engine belong traditionally to the most powerful tools of classical thermodynamics. Since the time of Bartoli [1] they are also used with appreciable success for the theoretical investigations of interaction between electromagnetic radiation and matter. Quite a crucial role in these studies plays the concept of adiabatic wall (partition) which is, as a rule, realized by means of an absolutely reflecting mirror made of a "perfectly conducting material". Application of such an abstraction to theoretical treatment of the properties of black-body radiation enclosed in a cavity enabled one to introduce the concepts of temperature and entropy of radiation into classical thermodynamics and, eventually, led to the derivation of the correct form for the dependence of the integral radiation density on temperature (Stefan-Boltzmann law). Among other results of these pioneering studies, an interesting theorem related to the subject of the present work should be mentioned [2], namely: "Expansion or compression of a cavity with adiabatic walls do not change the entropy of the radiation enclosed." It is a typical feature of this approach that the electromagnetic radiation was considered to be a self-contained entity, which might be only slightly influenced by the shape of the cavity and which was essentially independent of the quality of its walls. The very importance of the physical nature of the walls of the cavity on the processes involved was first realized only later by Planck [3], who simulated the physical properties of the walls by a finite set of abstract "oscillators". It was just his extensive research devoted to the black-body radiation which started the development of quantum mechanics and eventually led to the discovery of the so called zero-point (ZP) energy of his "oscillators". It was recognised later that such a zero-point energy is an intrinsic property of any physical system and now this concept plays the central role in modern theories describing e.g. the structure of quantum vacuum. Accordingly, there is a fluctuating electromagnetic field, existing quite independently of the source and thermal electromagnetic fields and persisting even in the absolute zero temperature limit where classically all motions cease [4,5]. It is further assumed that this "all-pervasive" electromagnetic radiation of unknown origin is homogeneous, isotropic and that its spectrum is invariant with respect to the Lorentz transformations (more generally, electromagnetic radiation is invariant with respect to the wider group of conformal mappings). It is interesting enough to notice that the latter property is decisive even for the analytical shape of the ZP electromagnetic spectrum. As was proposed by Boyer [6] on the basis of homogeneity arguments, there is only one possible form of spectral energy density $\rho(\omega)$ which is Lorentz invariant, namely:

$$\rho(\omega) = \hbar\omega^3 / 2\pi^2 c^3. \tag{1}$$

Serious disadvantage of this formula is obviously its divergence with respect to the integration over the infinite frequency range. To obtain from it a physically meaningful figures a rather laborious work with infinities [7] or an introduction of more or less arbitrary cut-off frequency [5] are required. Remarkably good prediction of ponderomotoric forces existing e.g. between two infinite perfectly conducting planes [8,9] (so called Casimir's effect) was obtained just by the computing of a finite difference of two infinite radiation forces acting on different sides of the said planes. In spite of these results, we consider such a reasoning unphysical. The corresponding mathematical procedure is, namely, based on an exact cancellation of semi-convergent infinite series containing the terms rather sensitive to the boundary conditions on the planes, which are, however, not well-defined [10]. Moreover, any macroscopic model of partitions involved in thermodynamic thought experiments with electromagnetic radiation should not a priori ignore their microscopic atomic structure without threatening serious danger of introducing a fatal error. It is evident that the formal manipulations with infinite number of terms describing high frequency electromagnetic modes, resulting just from the interaction of radiation with the partitions, is, in the case where the wavelengths are smaller than the inter-atomic distances, physically questionable and may provide the correct results only by chance. That is why we are convinced that the incorporation of intrinsic material properties of the reflecting or absorbing partitions into the thought experiments with radiation, at least to a certain degree of approximation, is an inevitable part of such considerations.

## Heat engine with cut-off frequency

In the present contribution we are trying to analyse a simple thermodynamic thought experiment performed with a heat engine containing ZP radiation, in which the reflecting walls are made of a material having an intrinsic upper cut-off frequency, more realistic, than a "perfect conductor" is.

The necessity of the existence of the upper cut-off frequency for the interaction of ZP electromagnetic radiation with ordinary matter can be explained as follows (cf. [5]). Setting aside the problem whether the electrodynamics can be extended to arbitrarily high frequencies or not (admitting e.g. without limits formula (1)), we discuss only common materials in which the response to the external electromagnetic radiation is mainly due to the electrons. It is obvious in this case that at very high frequencies $\omega \geq c/b$, where $b$ represents the extent of the structure and $c$ the velocity of light, the electrons are not able, because their speed is limited by $c$, to follow the electromagnetic vibrations and the field has to uncouple from them and the interaction ceases. Reasonable estimate for the upper frequency limit of such a decoupling is Compton's frequency $\omega_C = mc^2/\hbar$. Accordingly, it is assumed that all the parts of heat engine considered are made of a material which is, up to a certain cut-off frequency $\omega_K \leq \omega_C$, a perfect conductor (mirror) and is simultaneously fully transparent for frequencies $> \omega_K$. The engine itself consists of a cylindrical cavity provided with a piston which can move without friction. The position of the piston is measured by the distance $x$ between its inner side and the bottom of the cavity. The thought experiment is performed at $T = 0$ with an engine bathed in ZP radiation. Let us now expand the cavity by moving the piston with a constant velocity $V \ll c$ from its starting position at $x \approx c/\omega_K$. Because in this case $\omega_K$ is the gravest frequency there is at the beginning no other electromagnetic radiation except the isotropic ZP radiation of frequencies $> \omega_K$ freely penetrating through the walls of the cavity. The beams with frequencies belonging to a certain narrow band just above $\omega_K$, however, when meeting the inner side of the moving piston, will reflect from it, because their frequencies observed from moving coordinate system of the piston will be, due to Doppler's effect, smaller than $\omega_K$ (see Fig. 1a). For similar reasons, also the beams with frequency just below $\omega_K$ falling on the outer side of the moving piston will penetrate in the cavity (see Fig. 1b). All such reflected and penetrated beams will remain trapped, because their frequencies are smaller than the cut-off frequency $\omega_K$, within the cavity moving to-and-fro.

## Semi-quantitative relations

To be more specific, the frequency shift of a beam reflected from the inner side of moving mirror is given by the formula

$$\omega_1 = K\,\omega_0, \qquad (2)$$

where $\omega_0$ is the original frequency, $\omega_1$ the frequency after the first reflection, $K$ Doppler's factor which depends on velocity $V$ (or $\beta = V/c$) and incidence angle $\theta$. (For the piston moving outward from the cavity $K$ is evidently $< 1$.) Because reflection belongs to the group of conformal transformations, i.e. just to the group which preserves the spectral composition of ZP radiation, relation (2) should map any band of ZP spectrum (1) onto another band of the same curve. Thus any narrow band of ZP radiation lying just below $\omega_K$ can be transformed by a number of multiple reflections from the moving piston to the ZP radiation extended down to the gravest mode $\omega_G \approx c/x$ of the cavity. The process follows the formula:

$$\omega_N = K^N\,\omega_0, \qquad (3)$$

Where $\omega_N$ is the frequency of an original beam after $N$ reflections from the moving piston. To asses the limiting behaviour of the process just described some approximations are necessary. For example, for quasi-stationary displacement of the piston (i.e. $V \to 0$) the following estimate is valid:

$$K \approx (1 - 2\beta\cos\theta). \qquad (4)$$

At the distance $x$ between the piston and the reflecting bottom of the cavity the beam will suffer, during the time corresponding to the displacement of $dx$, $N$ reflections from the piston where $N \approx \cos\theta\,dx/(2\beta x)$. Consequently, with increasing $N$

$$\omega_N = \omega_0 (1 - 2\beta \cos\theta)^N \to \omega_0 \exp(-\cos^2\theta \, dx/x). \qquad (5)$$

From this formula it is obvious that the explicit dependence of the limiting frequency on velocity disappears and that for sufficiently large expansion $dx$ the cavity is filled practically down to zero frequency by ZP radiation. The process is, due to the time reversibility of beams, reversible also in thermodynamic sense. Particularly, for already filled large cavity ($x \gg c/\omega_K$, $\omega_G \to 0$) any change $dx \ll x$ is fully analogous to the classical reversible adiabatic process. Indeed, because the spectral composition corresponding to the distribution of ZP fluctuations given by (1) is in this case preserved during expansion or compression, the entropy change $dS=0$ as well as $dT=0$, which is fully in accordance with the definition of zero absolute temperature in classical thermodynamics [11].

## ZP radiation in a small cavity

The behaviour of small cavities (formally $x \approx c/\omega_K$) is, however, qualitatively different. The lower cut–off frequency corresponding to the ground mode of the cavity ($\omega_G \approx c/x$) reaches in this case high values comparable with $\omega_K$. The number of admissible modes within the interval ($\omega_G$, $\omega_K$) thus changes during the expansion or compression of the cavity appreciably (i.e. $dS \neq 0$) and the process can be no more treated as adiabatic. From this analysis it is apparent that partitions with upper cut-off frequency are equivalent to the classical adiabatic partitions only if the cavity is large enough, in other words, if $\omega_K \gg \omega_G$. Further, for $\omega \geq \omega_K$ the spectrum of ZP radiation, regardless of whether it is inside or outside of the engine, is not influenced by its presence. Therefore, for the computation of entropy changes and/or forces arising (i.e. Casimir forces) during the working cycle of our idealized engine, the only relevant are the electromagnetic modes from finite frequency band ($\omega_G$, $\omega_K$), where $\omega_G$ depends on the detailed geometry and absolute dimensions of the engine.

## Conclusions

Summarizing, we have introduced a first-step approximation model of reflecting partition with a sharp upper cut-off frequency $\omega_K$, which was used for the construction of an ideal heat engine working with ZP electromagnetic radiation.

Using semi-quantitative arguments a possible mechanism of filling real cavities with the ZP radiation based on Doppler's effect was suggested.

It has been shown that for large cavities ($x >> c/\omega_K$), in contrast to small cavities ($x \approx c/\omega_K$), the behaviour of partitions with the cut-off is essentially the same as that of adiabatic partitions known from classical thermodynamics. In other words, corresponding Carnot's process in an idealized heat engine is thus adiabatic (entropy change $dS = 0$) only if $x$ is sufficiently large and strongly non-adiabatic (entropy change $dS \neq 0$) in the case where $x$ is comparable with $c/\omega_K$.

We have further claimed that for considerations of ZP radiation in small real cavities the high frequency tail ($\omega > \omega_K$) is irrelevant. This statement is e.g. in apparent conflict with a standard approach to the computation of the Casimir forces in small systems with perfectly reflected (adiabatic) boundaries.

## Acknowledgement

This work has been supported by the Grant Agency of the Czech Republic under Contract No. 202/03/0410.

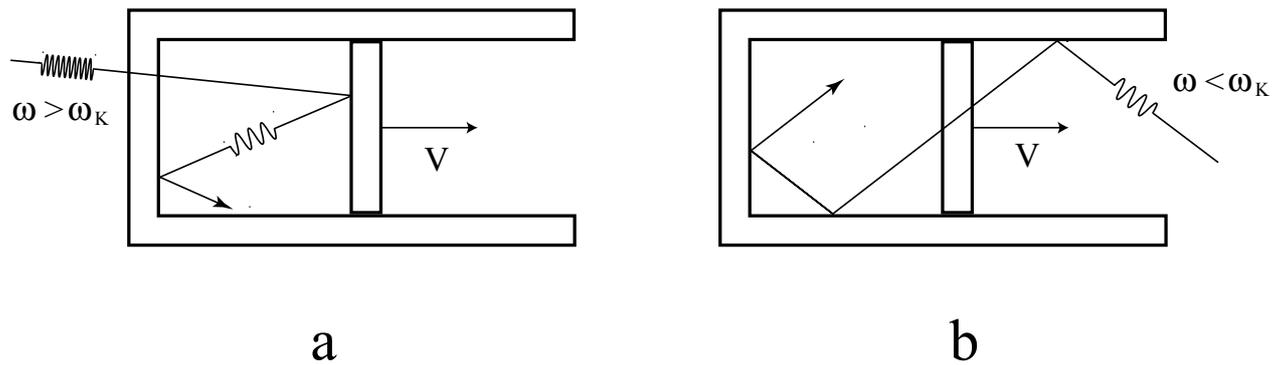

Fig. 1 : Two possible mechanisms of filling a cavity with ZP radiation based on Doppler's effect. (All the parts of the engine are made of mirrors with cut-off frequency $\omega_K$.)
a) The beam of frequency just above $\omega_K$ is reflected by a moving piston.
b) The beam of frequency just below $\omega_K$ penetrates the moving piston.